\documentclass[runningheads]{llncs}
\usepackage{paralist}
\usepackage{array}
\usepackage{booktabs}
\usepackage{multirow}
\usepackage{cite}

\usepackage[T1]{fontenc}
\usepackage{graphicx}
\usepackage{color}
\definecolor{citeblue}{rgb}{0.1,0,.4}
\usepackage[pdftex%
,colorlinks=true%
,bookmarks=false%
,linkcolor=citeblue%
,citecolor=citeblue%
,plainpages=false]{hyperref}

\usepackage{amsmath,amsfonts,amssymb}
\usepackage{xspace}
\newcommand{\R}{\mathbb{R}}
\newcommand{\SPEC}{\Phi_{\textrm{spec}}}
\newcommand{\SOL}{\Phi_{\textrm{sol}}}
\newcommand{\MIN}{\scalebox{0.7}[1.3]{$-$}}
\newcommand{\PLUS}{\raisebox{.15em}{\scalebox{0.7}[0.7]{$+$}}}
\newcommand{\vampire}{\textsc{Vampire}\xspace}

\newcommand{\cvc}{\textsc{cvc5}\xspace}

\usepackage{multicol}

\usepackage[misc]{ifsym}
\newcommand{\ENVELOPE}{\raisebox{.3em}{\scalebox{0.7}[0.7]{\Letter}}}

\begin{document}
\title{SMT and Functional Equation Solving\\ over the Reals: Challenges from the IMO}
\titlerunning{SMT and Functional Equation Solving over the Reals}
\author{%
     Chad E. Brown\inst{1}
\and Karel Chvalovsk\'y\inst{1}\orcidID{0000-0002-0541-3889}%
\and Mikol\'a\v{s} Janota\inst{1}\orcidID{0000-0003-3487-784X}\ENVELOPE%
\and Mirek Ol\v{s}\'ak\inst{3}\orcidID{0000-0002-9361-1921}%
\and Stefan Ratschan\inst{2}\orcidID{0000-0003-1710-1513}%
}

\authorrunning{Brown et al.}
\institute{%
Czech Technical University in Prague, CIIRC, Czechia\\
\email{mikolas.janota@cvut.cz}
\and Institute of Computer Science Academy of Sciences of the Czech Republic
\and University of Cambridge
}
\maketitle %
\begin{abstract}

We use SMT technology to address a class of problems involving uninterpreted
functions and nonlinear real arithmetic. In particular, we focus on problems
commonly found in mathematical competitions, such as the International
Mathematical Olympiad (IMO), where the task is to determine all solutions to
constraints on an uninterpreted function. Although these problems require only
high-school-level mathematics, state-of-the-art SMT solvers often struggle with
them. We propose several techniques to improve SMT performance in this setting.
\end{abstract}

\section{Introduction}%
\label{sec:introduction}

The \emph{AIMO Challenge}\footnote{\url{https://aimoprize.com/}} has garnered
significant interest from researchers, particularly within the machine learning
community, by focusing on mathematical olympiad problems~\cite{mathconstruct,bluff}. However, this
challenge is also highly relevant to the field of automated reasoning. In a
previous study~\cite{sc2}, the authors formally addressed the problem of
functional equations, which involves determining all functions that satisfy a
given set of equations. The problem is best illustrated by a simple example
where one is to find all $f\,:\,\R\rightarrow\R$, s.t.\

\begin{equation}\label{eq:example}
    \forall xy.f(x+y) = xf(y) + yf(x).
\end{equation}

Setting $y$ to $0$, yields $\forall x.\, f(x) = f(0)x$,
revealing that~$f$ must be linear and is entirely determined by the value
of $f(0)$. Furthermore, substituting~$x = 0$ reveals that~$f$ must
be identically zero, meaning the unique solution is $f(x) = 0$ (formally,
$f = \lambda x.\,0$). A solution should provide a clear description
of \emph{all} possible functions $f$, potentially parameterized by constants.

The initial approach of the authors~\cite{sc2} fixes a template (a polynomial with parameters
as coefficients) for the function $f$ and then determines how this template
should be parameterized to solve the given problem. The values of the
parameters are obtained by real quantifier elimination. Then, it is required to
prove that all possible solutions must fit within the template. To illustrate,
consider the ``linear'' template $f(x)=ax+b$ in equation~\eqref{eq:example} where
applying quantifier elimination on $\forall x$ gives all the possible values of
$a$ and $b$, which is simply $a=b=0$. In the second phase, it needs to be proven that all
possible $f$ must be linear, which can be stated as follows.
\begin{equation}\label{eq:example:linear}
    (\forall xy.f(x+y) = xf(y) + yf(x)) \Rightarrow (\exists a,b\forall x.f(x)=ax+b)
\end{equation}

It turns out that proving such an implication is the bottleneck of the
approach. That is, even if it is easy to find every~$f$ within the template
that solves the equality, proving that the template is sufficient to cover all
the solutions, is hard for state-of-the-art solvers for Satisfiability Modulo Theories (SMT).

The problem under investigation presents a dual challenge: function synthesis
and comprehensive coverage verification (``there are no more $f$'s''). Function synthesis alone represents a
significant computational challenge with numerous methodological
approaches~\cite{sygus,hozzova-cade23,hozzova-ijcar24,ratschan-mfcs23,bradley-vmcai06,ge-moura-cav09},
which we address through a template-based methodology~\cite{srivastava13}
coupled with quantifier
elimination~\cite{Tarski:51,Davenport:88,Weispfenning:88}. The principal
contribution of this paper, however, lies in establishing the comprehensive coverage verification.
We introduce novel inference techniques that
enhance solvers' performance by generating auxiliary lemmas and strategic
quantifier instantiations prior to invoking SMT solvers. These techniques
generalize to other challenging problems in real arithmetic that, despite their
compact representation, present considerable difficulty in solving.

\noindent The key contributions of this paper are as follows:
\begin{compactitem}
    \item Development of an exploratory procedure that discovers novel lemmas to facilitate proof construction.
    \item Formulation of specialized instantiation-based techniques designed for computationally intensive quantified problems.
    \item Implementation of these advances in a unified framework that leverages state-of-the-art SMT solvers.
    \item Introduction of a benchmark suite to evaluate these techniques,
    demonstrating meaningful performance improvements over  the existing
  method.
\end{compactitem}

\section{Preliminaries}%
\label{sec:preliminaries}

Throughout the paper, a basic understanding of first-order logic is
expected~\cite{handbookAR}. We use SMT
solvers~\cite{Barrett2018} mainly in a black-box fashion, where the primary target
is the combination of the theories of nonlinear real arithmetic and uninterpreted
functions (\texttt{UFNRA}).

By a \emph{substitution} in a formula~$\phi$ we mean a mapping from all free
variables of $\phi$ to terms and its application is denoted as $\phi[x_1\mapsto
t_1,\dots,x_n\mapsto t_n]$ where all $x_i$ are replaced by the corresponding
term $t_i$ simultaneously. For a vector of variables $\vec{x}$, a quantified
formula $\forall\vec{x}.\,\phi$, and a substitution $s$, an
\emph{instantiation} is the formula~$\forall\vec{x}'.\,\phi[s]$, where
$\vec{x'}$ are the free variables introduced by~$s$.

SMT solvers attempt to solve quantified problems by adding quantifier
instantiations.%
\footnote{If formulas are not prenex, new instantiations are added in the form
  of the implication $(\forall\vec{x}.\,\phi)\rightarrow\phi[s]$,
  but through this paper we assume prenex form.
}
These instantiations only introduce ground
terms so that they can be sent to the ground subsolver.
There is a large body of research on quantifier
instantiation in SMT, with syntactic-driven approaches
(\emph{e-matching}~\cite{DetlefsNS05} or syntax-guided
instantiation~\cite{niemetz-tacas21}), semantic-driven
(\emph{model-based}~\cite{ge-moura-cav09,reynolds-cade13}),
\emph{conflict-based}~\cite{reynolds-fmcad14}, and \textit{enumerative
instantiation}~\cite{janota2021fair,ReynoldsBF18}.
Our experiments also use the automated theorem prover
\vampire~\cite{vampire} as an SMT solver. \vampire does not explicitly instantiate but relies
on superposition~\cite{handbookAR}.

\emph{Quantifier Elimination (QE).}  We need to be able to handle nonlinear
arithmetic over the real numbers. The corresponding logical theory---the theory
of real closed fields---allows QE~\cite{Tarski:51}. This
means, that there is an algorithm that takes any formula with the signature $\{
0, 1, +, \times, =, \leq\}$ (in practice, further symbols that can be
expressed in terms of these) and computes a quantifier-free formula that is
equivalent w.r.t.\ the axioms of the theory of real closed fields. QE
in the theory of real-closed fields is highly costly, both in
theory~\cite{Weispfenning:88,Davenport:88} and in practice. Hence it is usually
beneficial, and often indispensable, to exploit the specific structure of the
quantifier elimination problems at hand.
For QE we use QEPCAD~\cite{CollinsH91,Brown:04} via the tool Tarski~\cite{tarski}. Additional techniques are discussed in
Section~\ref{sec:qeeqs}.

\section{Problem Statement and Templates Revisited}\label{sec:templates_revisited}
In many cases, the problem is given as an equation, just as in
example~\eqref{eq:example}, but more complicated problem descriptions may
appear. For instance, apart from the equation there might also exist the
requirement that the function $f$ must be monotone, injective, etc.
Similarly, the solution to a problem might be a single function but also more
complicated classes of functions often appear, e.g.\ identity with a positive
shift, which can be described as~$\exists c.\,c>0\land \forall
x.\,f(x)=x+c$. We formalize this in the following.

The \emph{input} to a problem is a \emph{specification} of a function~$f\,:\,\R\rightarrow\R$, which is a
sentence~$\SPEC$ in the language of first-order real arithmetic augmented with a
unary function symbol~$f$.\footnote{%
  Competitions may contain problems with multiple functions and with different
  domains, e.g.\ $\mathbb{Q}$. These we currently do not support. We do support
  certain restrictions on the reals, e.g.\ $\R^+$ but we avoid discussing this
  here for the purpose of conciseness.}

A \emph{solution} to a problem is a specification $\SOL$ that is logically equivalent to the specification~$\SPEC$
but in \emph{solved form}, defined as follows. A sentence of the form
$\exists \vec{a}.\Gamma\land\forall x.f(x)=t$ is in \emph{solved form} iff $f$
occurs in neither $\Gamma$ nor $t$. We consider $\exists \vec{a}.\forall
x.f(x)=t$ to be the same as the special case where $\Gamma$ is $\top$. Note that
the solution might also be the empty class of functions, in which case~$\Gamma=\bot$. A disjunction $\Psi_1\lor\Psi_2$ is
in \emph{solved form} if each disjunct is.

In Example~\eqref{eq:example}, the given specification $\SPEC$ is $\forall
xy.f(x+y) = xf(y) + yf(x)$. The corresponding solved form $\SOL$ is $\forall
x.f(x)=0$.

\emph{Finding a solution} to a specification $\SPEC$ is done in two phases: First
construct a candidate solution $\SOL$, and then attempt to prove
$\SPEC\Leftrightarrow \SOL$. If this fails, potentially look for another
candidate solution.

When proving $\SPEC\Leftrightarrow \SOL$, we focus on the implication
$\SPEC\Rightarrow\SOL$, i.e.\ that the class of functions described by the
solution $\SOL$ covers all functions specified by~$\SPEC$. Typically, this
implications is significantly harder to prove and the converse follows by the
construction of~$\SOL$.

In the context of an SMT problem, proving $\SPEC\Rightarrow\SOL$ corresponds to
assuming $\SPEC\land\lnot\SOL$ and proving unsatisfiability. We typically start
with a skolemized version of $\lnot\SOL$, introducing new constants.

Following the initial work~\cite{sc2}, we look for solutions in the quadratic
polynomial template $ax^2+bx+c$. To obtain all possible solutions within the
template, we first plug the template into the specification formula and then
eliminate all quantified variables using QE. This results in a formula
containing only~$a, b, c$, from which the solved form is obtained heuristically,
see~\cite[Sec.3.3]{sc2}.

Note that the coefficients appear $a,b,c$ appear quantified existentially in the
solution as per solved form. However, since we are dealing with a polynomial
template, they can be eliminated by setting $c=f(0)$, $a=(f(1)+f(-1))/2-c$, $b=f(1)-a-c$,
see~\cite[Sec.~3.1]{sc2} and \emph{Lagrange
interpolation}~\cite{burden2015numerical}. For example,
$\exists c.\,c>0\land \forall x.\,f(x)=x+c$ is simplified to
$f(0)>0\land \forall x.\,f(x)=x+f(0)$.

To reduce the computational cost of QE, we also use variations of
the quadrat\-ic template by setting some of the coefficients to~$0$ (linear
$bx+c$, quadratic monomial $ax^2$, constant $c$), see~\cite[Sec.~3.1]{sc2}.
An alternative, specialized technique that improves QE for equations is described in
Section~\ref{sec:qeeqs}.

At this stage, all solutions of the specification are within the template.
But there may exist other solutions, \emph{outside} of the template.
The earlier approach~\cite{sc2} tried to prove that any $f$ satisfying the
specification must follow the template. We have, however, observed that it is
more efficient to first find the solution within the template and then prove
that there are no other solutions.

Coming back to the introductory example~\eqref{eq:example}, rather than proving
that the solution must be linear, we first identify the solution
$f(x)=0$, and then prove that this is the only solution possible, leading to
the following SMT problem.

\begin{multicols}{2}
\noindent\begin{equation}
  \forall xy.f(x+y) = xf(y) + yf(x)
\end{equation}
\noindent\begin{equation}
  \lnot(\forall x.f(x)=0)\label{eq:absence}
\end{equation}
\end{multicols}

We surmise that the reason why these proofs are easier for the SMT solver is
that there are constants from the proposed solution.
As noted above, the formula~\eqref{eq:absence} is skolemized as $f(c)\neq 0$,
with $c$ being a fresh constant. While we could rely on the SMT solver to
perform the skolemization, we explicitly perform it in our implementation,
since the skolem constant is useful for our techniques.

\subsection{Quantifier Elimination on Equalities}\label{sec:qeeqs}%

The high cost of quantifier elimination in the theory of real-closed
fields warrants the exploitation of the specific structure of the
problems we target. In our case, we have problems that are
the result of substituting a template for the function symbols in the given
specification formula. In many problems, the specification formula is a
functional equation~\cite{Kuczma2009}. Applying the template to it yields a
formula of the form $\forall\vec{x}.\: P(\vec{a},\vec{x})=0 \wedge F$, where $P$
is a polynomial, $\vec{a}$ stands for the parameters of the template, and
$\vec{x}$ stands for the universally quantified variables of the original
functional equation.
The part $F$ may come from additional requirements on the function, e.g.\ $f$ is
non-decreasing.

Now observe that the only polynomial that is zero everywhere is the zero
polynomial. Hence, denoting $P(\vec{a},\vec{x})$ as a polynomial in
$\vec{x}$ with parametric coefficients $p_1(\vec{a}),\dots,p_k(\vec{a})$, we
can rewrite the problem to the form \[ p_1(\vec{a})=0\wedge\dots \wedge
p_k(\vec{a})=0\wedge \forall \vec{x}. F, \] which is usually considerably
easier to solve. Moreover, in the special case where $F$ is not present, this
already describes the result of quantifier elimination. For example, substituting the template $ax^2+bx+c$ into Equation~\eqref{eq:example}, one arrives at a polynomial equation in the variables $a,b,c,x,y$ that contains (among others) the monomials $ax^2, (b-c)x, c$, resulting in a system of equations that contains (again among others) the equations $a=0, b-c=0, c=0$, from which one can immediately read off the solution $a=0, b=0, c=0$. 

In our computations, we use the SymPy
package~\cite{sympy}  to obtain the concrete values of the
coefficients.

\section{Quantifier Instantiation}\label{sec:inst}

Our initial experiments indicate, that the reason why problems are not being
solved, is the difficulty of finding the right quantifier instantiations.
In general, the ground part could potentially also be difficult because it requires the
theory of nonlinear real arithmetic, but by manually testing with the
appropriate quantifier instantiations, we have observed that the SMT solvers
are typically successful. SMT instantiation techniques are
tailored to work fast within the DPLL(T) framework, which explains why they have
less success on problems that are small but highly challenging.
In this section we introduce instantiation techniques targeting such problems.

Our experiments also use the automated theorem prover \vampire, which
does not explicitly instantiate during proving but also benefits from our
techniques because we explicitly add instantiations into the input formula.

\subsection{Partial Instantiations with Simple Terms}\label{sec:partial}%
As seen in our introductory example~\eqref{eq:example}, valuable information can
be obtained by substituting concrete values, such as 0 and 1, which will likely
simplify the problem.
Even though cvc5 has the enumerative mode~\cite{janota2021fair}, it only relies
on the ground terms in the formula---and 0,1 might not be in it---and furthermore,
it always grounds \emph{all} variables under a quantifier. However, partial
instantiations, where some variables are left as the original variables, provide
powerful facts (recall that substituting 0 for $x$ in~\eqref{eq:example}
immediately shows that $f$ must be linear). It may also be helpful to
instantiate with other small ground terms, e.g., skolem constants. We
illustrate this on the problem \textbf{U10} from~\cite{musil}:
\begin{equation}
  f(x^2+y) + f(f(x) - y) = 2f^2(x) + 2y^2.\label{eqn:c4}
\end{equation}
For our purposes, here we assume the solution $f(x)=x^2$
has been found and it only remains to prove there are no more solutions.
As an SMT problem, this means we assume the identity~\eqref{eqn:c4}
and assume $f(c)\not=c^2$ for a constant $c$.
The resulting problem is unsatisfiable, although \cvc and \vampire
are unable to determine unsatisfiability with a 10~minute time limit.

It is enough to instantiate $(x,y)$ in~\eqref{eqn:c4}
with six relatively small instantiations:
$(0,0)$, $(0,f(0))$, $(0,-c^2)$, $(c,0)$, $(c,f(c))$ and $(c,-c^2)$.
Let us briefly consider why these instantiations are enough. %
The instantiations $(0,0)$ and $(0,f(0))$ easily yield $f(0)=0$.
The instantiation $(c,f(c))$ yields $f(c^2+f(c)) = 2f^2(c) + 2(f(c))^2$
while $(c,-c^2)$ yields $f(f(c)+c^2)=2f^2(c) + 2c^4$,
which together give $(f(c))^2 = c^4$.
Since we know $f(c)\not=c^2$, we must have $f(c)=-c^2$.
Combining this with $f(f(c)+c^2)=2f^2(c) + 2c^4$
we have $f^2(c) = -c^4$.
The instantiation $(c,0)$ yields $f(c^2) = f^2(c)$, hence $f(c^2) = -c^4$
and $f(c^2) = f(f(c)) = f(-c^2)$.
The instantiation $(0,-c^2)$ yields $f(-c^2) + f(c^2) = 2c^4$
and so $-2c^4 = 2c^4$ giving $c=0$. However, $c=0$ contradicts $f(0)=0$
and $f(c)\not=c^2$.

If we add these six instantiations manually, both \cvc and \vampire
refute the problem within 1~second.
The largest term used in the instantiations above is $-c^2$ (or, more properly, $-(c\cdot c)$).
An early partially successful experiment
was to enumerate all terms up to the size of $-c^2$ and try selected pairs of instantiations
using these terms.
However, it proved more successful to use \emph{partial} instantiations.
Note that in each of the 6 pairs above the instantiation for $x$ is simply $0$ or $c$.
This led us to try instantiating each variable
with a very small term (e.g., $0$ and $c$) and leave the other variable free.
This yields four partially instantiated versions of~\eqref{eqn:c4}.
We also have the option of including the original equation~\eqref{eqn:c4} with both variables free.
If we include only the four partially instantiated equations,
\vampire (using a portfolio) refutes the resulting problem in less than 3 seconds.
The specific strategy in the portfolio that solves the problem takes 0.2 seconds.
An inspection of the proof generated by \vampire demonstrates that
it is using instantiations beyond the six pairs described above.
In particular, \vampire effectively uses the pair $(c,-f(-c))$ for $(x,y)$
at one point of the proof,
which is not one of the six pairs with which we started. %

Our implementation considers a slightly larger set of small terms for partial instantiations,
with the minimal set being $0$, $1$ and all skolems in the problem (e.g., $c$ in \textbf{U10})
and the maximum set additionally including other ground terms occurring in the equation
(e.g., $2$ in \textbf{U10}).
When using $0$, $1$ and $c$ for the partial instantiations, \vampire
(using the strategy that solves the problem with $0$ and $c$) can still
find a refutation, but in 0.44 seconds.
When using $0$, $1$, $2$ and $c$ for the partial instantiations, \vampire
(with the same strategy) solves the problem in 2.3 seconds.

\subsection{Theory-Unification by Equation Solving}%
\label{sec:nice:eqs}
E-matching is an SMT technique for finding instantiations that results in terms that already
appear in the formula modulo equality~\cite{DetlefsNS05}. For our purposes, equality
alone is insufficient as we also need \emph{theory reasoning}. In particular,
we would like matching to be aware of real arithmetic. This is illustrated by the following example
(\'Uloha~6 by Musil~\cite{musil}). Consider the following specification for~$f$.

\begin{equation}
  \forall xy.\ f(x+y)-f(x-y)=xy\label{u6:spec}
\end{equation}

The quadratic template gives the candidate solution $f(x)=x^2/4+f(0)$,
which is negated and skolemized, resulting in the following.

\begin{equation}
  f(c)\neq c^2/4+f(0).\label{u6:sol}
\end{equation}

To prove that all possible $f$ are covered, unsatisfiability of~\eqref{u6:spec}$\land$\eqref{u6:sol} needs to be shown.
This is done by substituting both $x$~and~$y$ with $c/2$ in~\eqref{u6:spec}:
\begin{align}
  f\left(c/2+c/2\right) -
  f\left(c/2-c/2\right) =c^2/4\label{half} \\
  f(c) = c^2/4+ f(0) \label{eq:c2}&
\end{align}

Traditional E-matching cannot discover this substitution because $c/2$
does not occur in the existing formula. One could try to factor the
term $c^2/4$, but we take a different approach. We observe it is often useful
to derive instantiations where $f$ is simply applied to $x$. Ideally, we obtain a
definition-like equality, i.e.\ $f(x)=t$, where $t$ does not contain $f$, or
$f$ is only applied on ground terms.

We look for such instantiations by the following heuristic. Collect
all the arguments $A$ to $f$ in a quantified subexpression---consider only top-level
arguments in the form of a polynomial.
Non-deterministically partition $A$ into $A_z$ and $A_0$. Find a substitution
that makes all the arguments in $A_z$ equal to a fresh variable $z$ and sets
all arguments in $A_0$ to $0$. Add all instantiations by the obtained
substitutions. The Python package SymPy~\cite{sympy} is used to solve the
equations.

In example~\eqref{u6:spec}, considering all possible subsets of $\{x+y,x-y\}$
yields the following equations and their corresponding solutions

\[
  \begin{array}{lcr}
  x+y=z, x-y=0 &\quad\cdots\quad  &\quad  x\mapsto z/2, y\mapsto z/2\\
  x+y=0, x-y=z &\quad\cdots\quad  &\quad  x\mapsto z/2, y\mapsto -z/2\\
  x+y=z, x-y=z &\quad\cdots\quad  &\quad  x\mapsto z, y\mapsto 0 \\
  x+y=0, x-y=0 &\quad\cdots\quad  &\quad  x\mapsto 0, y\mapsto 0
  \end{array}
\]

The first substitution yields, after simplification, $\forall
z.\,f(z)-f(0)=z^2/4$. Then the underlying SMT solver just needs to substitute
$c$ for~$z$ to obtain a contradiction as in~\eqref{eq:c2}.

Note that setting arguments in $A_0$ to $0$ is a somewhat arbitrary choice, but
in our benchmarks, $0$ typically leads to further simplification in the formula.
An alternative would be to substitute a fresh constant~$k$ and solve for it as
well.

\section{Lemma Generation}%
\label{sec:newfacts}

Coming up with lemmas that improve the solvers' performance is a notoriously
challenging task, but we propose to look for useful conjectures by generating
relevant ground instances of simple equations. For example,
in~\eqref{eq:example}, it holds that $f(0)=0$, $f(1)=0$, $f(0)=f(1)$, etc. We
generate such simple statements (essentially by brute-force) and test whether
they hold in the given formula to be solved. If they do, we add them to the
problem, as an additional axiom.

The equations are generated systematically from terms of small depths. We start
with a small set of initial terms $0$, $1$, skolem constants, and possible other
numbers occurring in the problem. These are subsequently combined using
addition, subtraction, multiplication, and the application of the function~$f$.

Since this systematic process quickly explodes, we try to limit the conjectures
being generated by making substitution into the candidate solution.
For example, to show that $f(x)=x\vee f(x)=-x$ (two
solutions), we substitute a skolem $c$, an initial
term, for $x$, and obtain $f(c)=c\vee f(c)=-c$ as a conjecture. This
makes it possible to express more general conjectures for problems
with multiple solutions. Moreover, we also add the individual
equations as conjectures, e.g.\ $f(c)=c$ and $f(c)=-c$ to obtain
them sooner.

We use the previous techniques to iteratively generate
conjectures. Conjectures that are proven are then added to our original
SMT~problem as lemmas and we attempt to solve the SMT~problem again. If we
fail, we restart the conjecture-generating process from scratch. This
has various advantages. First, as we use a portfolio of solvers, they
can exchange these proven lemmas. Second, the derived lemmas influence
quantifier instantiations. Moreover, we prune conjectures for
redundancies by quickly checking whether they follow from the
previously derived lemmas alone. Note that these quick checks involve
only ground formulas.

\section{Experiments}%
The implementation and data is available at~\cite{data}.
For QE we use Tarski~\cite{tarski} due its capability to read SMT\@.
Alternatively, one could also use Z3's QE functionality but we had a better
experience with Tarski overall.
For SMT queries several solvers and their configurations are used, run in parallel and stopping all, once one finishes.
We use Z3~\cite{z3}\footnote{version 4.12.1, with default options and \texttt{-memory:32768}},
\vampire~\cite{vampire}\footnote{%
  version 4.8 linked with Z3 4.9.1.0 and options \texttt{--input\_syntax smtlib2 --mode portfolio --schedule smtcomp\_2018 --cores 1}},
and \cvc~\cite{cvc5}\footnote{version 1.1.3-dev.72.2b4ca00c2} in several configurations:
$\{\texttt{enum-inst}\}$,
$\{\texttt{no-e-matching}, \texttt{enum-inst}\}$,
$\{\texttt{simplification=none}, \allowbreak \texttt{enum-inst}\}$,
$\{\texttt{mbqi}\}$,
$\{\texttt{no-e-matching}, \texttt{no-cbqi}, \texttt{enum-inst}\}$.
The experiments were performed on machines with two AMD EPYC 7513 32-Core
processors and with 514~GiB~RAM with 10 problem instances run in parallel.
Each problem instance has the CPU time limit 3600s and calls up to 8 SMT solvers in parallel. Each SMT solver has the time limit 120s (5s for lemmas).

As in~\cite{sc2}, we use a problem collection by Musil~\cite{musil} comprising
79 problem instances from several sources.
  We collected more problems from competitions. We scraped the problems
  from Art Of Problem Solving (AoPS)~\cite{AOPS} for
  International, Regional, and National contests. Then we heuristically
  filtered problems resembling functional equations, obtaining 754 problems
  as text. Then we translated these problems to our custom language for
  functional equations, which we converted into SMT\@. This procedure
  successfully produced 343 problems.

To evaluate the techniques we have performed an ablation study of the proposed
approaches. A summary of the results can be found in
Table~\ref{tab:experiments}.
The default approach is to first run SMT queries with additional instances obtained by Theory-Unification (TU), see Section~\ref{sec:nice:eqs}. This part corresponds to {\MIN}PI{\MIN}L in Table~\ref{tab:experiments}. If this fails, we produce more instances using partial instantiations (PI), see Section~\ref{sec:partial}, and query SMT solvers again. The problems solved by these two steps correspond to {\MIN}L in Table~\ref{tab:experiments}. If this also fails, we run the lemma generation loop (L) interleaved with SMT queries when new lemmas are proven, see Section~\ref{sec:newfacts}, until the time limit. 
We also consider the \emph{virtual best solver (VBS)},
which considers the shortest solving time for each instance across all
configurations of our tool.
The results show that all the techniques contribute to the default mode. Lemmas
and partial generation have the greatest impact---turning them off causes a loss of
half of the instances.

\begin{table}[tb]
  \centering
\caption{The number of solved problems. The default (Def.)\ uses EQ, PI, TU and L\@. In {\MIN}EQ the original equation is removed during PI\@. {\PLUS}FI means partial instantiations of equation by simple terms and terms
  obtained by one application of $+$,$-$,$\times$, or $f$ from them
  into up to three variables. {\MIN}PI means no partial
  instantiations (Sec~\ref{sec:partial}) by simple terms into one variable. {\MIN}TU means no Theory-Unification (Sec~\ref{sec:nice:eqs})\@. {\MIN}L means
  no lemmas (Sec~\ref{sec:newfacts}). Base is {\MIN}PI{\MIN}TU{\MIN}L\@. VBS is the virtual best solver.}
\label{tab:experiments}
  \sf
\newcolumntype{P}[1]{>{\centering\arraybackslash}p{#1}}
\begin{tabular}{llP{.05\textwidth}P{.06\textwidth}P{.1\textwidth}P{.055\textwidth}P{.055\textwidth}P{.1\textwidth}P{.05\textwidth}P{.08\textwidth}P{.08\textwidth}P{.07\textwidth}P{.05\textwidth}}
  \toprule
  & & Def. & {\MIN}EQ & {\MIN}EQ{\PLUS}FI & {\MIN}PI & {\MIN}TU & {\MIN}PI{\MIN}TU & {\MIN}L &{\MIN}PI{\MIN}L &{\MIN}TU{\MIN}L & Base & VBS \\ \midrule
  {\multirow{2}{*}{Musil} }& total & 21 & 21 & 20 & 18 & 17 & 13 & 19 & 17 & 14 & 13 & 22 \\
                    & unique & 0 & 0 & 1 & 0 & 0 & 0 & 0 & 0 & 0 & 0 & \\  
  {\multirow{2}{*}{AoPS} } & total &  77 & 75 & 64 & 60 & 76 & 47 & 64 & 39 & 58 & 33 & 87 \\
                    & unique &  2 & 1 & 5 & 2 & 1 & 0 & 0 & 0 & 0 & 0 &\\  \midrule
  \bottomrule
\end{tabular}
 \end{table}

\subsection{Example Solved Problem}

To demonstrate the current capabilities, we show an example of a problem that the system could solve, contrary to its predecessor. We also show the standard human solution to give a sense of difficulty. It is problem 19 of the 2005 Postal Coaching contest in India. In our benchmark, it can be found under the AoPS code \texttt{c1068820h2554984p21862259}.

Find all functions $f : \mathbb{R} \mapsto \mathbb{R}$ such that $f(xy+f(x)) = xf(y) +f(x)$ for all $x,y \in \mathbb{R}$. Let $(a,b)$ denote plugging $x=a$ and $y=b$ into the original equation.
To solve the problem, we start with two substitutions.

\[
\vcenter{\halign{%
\hfil$\displaystyle# : $\quad&\hfil$\displaystyle# = {}$&$\displaystyle#$\hfil\cr
(0,x) & f(f(0)) & f(0), \cr
(f(0), 0) & f(f(f(0))) & f(0)^2 + f(0). \cr
}}
\]

By combining these two equations, we obtain $f(0)^2 = 0$, therefore $f(0) = 0$. Now, we plug $(x,0)$ into the original equation and obtain $f(f(x)) = f(x)$.
We continue with two symmetric substitutions
\[
\vcenter{\halign{%
\hfil$\displaystyle# : $\quad&\hfil$\displaystyle# = {}$&$\displaystyle#$\hfil\cr
(x,f(x)) & f(xf(x) + f(x)) & xf(x) + f(x), \cr
(f(x),x) & f(f(x)x + f(x)) & f(x)^2 + f(x). \cr
}}
\]
Notice that the left hand sides are equal, and equating the right hand sides simplifies to $f(x)^2 = xf(x)$, equivalently $f(x) = 0$, or $f(x) = x$.

In theory, for any real number $x$, there could be an independent choice of whether $f(x) = x$ or $f(x) = 0$. Now, we will prove that in fact, across all the values of $x$, the choice must be the same.

For contradiction, assume that there are two nonzero numbers $a,b$ such that $f(a) = a$ and $f(b) = 0$. Let us plug $(a,b)$ into the equation:
$f(ab + a) = a$.
Because $a\neq 0$, our characterization of possible values for $f(x)$ forces $a = ab+a$. However, this equation simplifies to $ab = 0$, and cannot be satisfied by nonzero values.
Therefore, the only two solutions are $f(x) = 0$ for all $x\in\mathbb{R}$, and $f(x) = x$ for all $x\in\mathbb{R}$. We easily check that these two functions satisfy the given equation.

Our solver automatically solves the same problem as follows.
First, using the template and quantifier elimination it is easy to obtain
the two solutions, and as usual it remains to prove there are no other solutions.
That is, the solver needs to prove unsatisfiability of the identity
$f(xy+f(x))=xf(y)+f(x)$ along with the two skolemized
disequations $f(c_1)\neq c_1$ and $f(c_2)\neq 0$.
Using the techniques described in this paper, the solver additionally adds
partial instantiations of $f(xy+f(x))=xf(y)+f(x)$.
The solver then generates and proves lemmas.
First, it derives lemmas $f(0) = 0$ and $f(1)=1\vee f(1)=0$.
Second, it derives $f(c_1)=0$, $f(c_2)=c_2$ and $f(1)=1$ using the previously derived lemmas.
Each lemma was proven using \vampire.
Using all these lemmas, the unsatisfiability of the whole set can be shown instantly by all solvers in our portfolio but one.\footnote{The one is \cvc called with enumeration and no simplification.}
Note that our informal proof above shows that if $f(a) = a$ and $f(b)=0$, then either $a$ or $b$ must be $0$.
By the same argument, the automatically derived lemmas $f(c_1)=0$ and $f(c_2)=c_2$ imply either
$c_1$ or $c_2$ must be $0$, contradicting $f(0)=0$.

\section{Conclusions and Future Work}%
\label{sec:conclusion}

We develop a pipeline that solves for all real functions
satisfying a specification $\SPEC$. We first use quantifier elimination in order
to solve for all quadratic solutions to the specification, giving a candidate
solution $\SOL$ satisfying $\SOL\Rightarrow\SPEC$. We then try to prove
$\SPEC\Rightarrow\SOL$ in order to conclude $\SOL$ exhausts the class of all
solutions. Such implications are challenging for SMT solvers and we have
combined several novel techniques (and several SMT solvers) to automatically
justify the implication.

One of the techniques is to instantiate some quantifiers before calling the SMT
solver. We look for instantiations by solving equations (over the theory
of the reals) and sometimes simply use small instantiations. Apart from
instantiations, we also generate and prove potential lemmas and add them to the
problem. This approach enables synergy between SMT solvers since
one solver may prove a lemma and another solver uses the lemma in the final
proof.

The introduced techniques significantly contribute to the number of problems
solved, which suggests that such techniques may be useful in a more general SMT
setting. Proving that there are no more solutions is also of interest in
automated synthesis, since more (unknown) solutions indicate ambiguity in
the specification---ambiguity has been studied by
Kun\v{c}ak~et~al.~\cite{DBLP:conf/pldi/KuncakMPS10} in the context of synthesis
but only for specifications that admit deskolemizing~$f$ (meaning that $f$ is
always applied to the same arguments).

We found some helpful instantiations by solving equations. A more general
version of this technique would be to use \emph{rigid E-unification} to solve
equations in the theory of the reals along with the assumed equations for the
unary function $f$~\cite{BackemanRummer-frocos2015}. Some limited support for
linear arithmetic in E-matching has been developed by Hoenicke and
Schindler~\cite{HoenickeSchindler-vmcai2021}.
We leave this for future work.
Unification with abstraction~\cite{reger-tacas18} can help Vampire
obtain useful instantiations for clauses with theory and nontheory parts (e.g., some
parts purely involving reals and other parts involving the
uninterpreted function symbol $f$).
Updating to use more recent work on
unification with abstraction~\cite{bhayat-lpar23} may be a fruitful avenue of future work.
Another topic for future work is
to improve the synthesis of potential lemmas in order to produce
lemmas beyond disjunctions of ground equations.

Another interesting topic for future research is how to handle richer templates.
For example, we could consider the quotient of two polynomials (of bounded degree).
Having a polynomial in the denominator introduces issues with how division by zero
is handled (or avoided).

\begin{credits}
\subsubsection{\ackname}
The research was
supported by the Ministry of Education, Youth and Sports
within the dedicated program ERC~CZ under the project \textsf{POSTMAN} no.~LL1902,
by the Czech Science Foundation grant no.~25-17929X, and
by the European Union under the project \textsf{ROBOPROX}
(reg.~no.~CZ.02.01.01/00/22\_008/0004590).
This article is part of the \textsf{RICAIP} project that has received funding from
the European Union's Horizon~2020 research and innovation programme under grant
agreement No~857306. Stefan Ratschan's work was supported by the research
programme of the Strategy AV21 AI: Artificial Intelligence for Science and
Society and institutional support RVO:67985807.

\subsubsection{\discintname}
The authors have no competing interests to declare that are
relevant to the content of this article.
\end{credits}

\bibliographystyle{splncs04}

\begin{thebibliography}{10}
\providecommand{\url}[1]{\texttt{#1}}
\providecommand{\urlprefix}{URL }
\providecommand{\doi}[1]{https://doi.org/#1}

\bibitem{AOPS}
Art of problem solving, \url{https://artofproblemsolving.com/community/c13_contest_collections}

\bibitem{sygus}
Alur, R., Bodik, R., Juniwal, G., Martin, M.M.K., Raghothaman, M., Seshia, S.A., Singh, R., Solar-Lezama, A., Torlak, E., Udupa, A.: Syntax-guided synthesis. In: 2013 Formal Methods in Computer-Aided Design. pp.~1--8 (2013). \doi{10.1109/FMCAD.2013.6679385}

\bibitem{BackemanRummer-frocos2015}
Backeman, P., R{\"{u}}mmer, P.: Free variables and theories: Revisiting rigid {E}-unification. In: Lutz, C., Ranise, S. (eds.) Frontiers of Combining Systems - 10th International Symposium, FroCoS 2015, Wroclaw, Poland, September 21-24, 2015. Proceedings. Lecture Notes in Computer Science, vol.~9322, pp. 3--13. Springer (2015). \doi{10.1007/978-3-319-24246-0\_1}

\bibitem{cvc5}
Barbosa, H., Barrett, C.W., Brain, M., Kremer, G., Lachnitt, H., Mann, M., Mohamed, A., Mohamed, M., Niemetz, A., N{\"{o}}tzli, A., Ozdemir, A., Preiner, M., Reynolds, A., Sheng, Y., Tinelli, C., Zohar, Y.: cvc5: {A} versatile and industrial-strength {SMT} solver. In: Tools and Algorithms for the Construction and Analysis of Systems, {TACAS}. LNCS, vol. 13243, pp. 415--442. Springer (2022). \doi{10.1007/978-3-030-99524-9\_24}

\bibitem{Barrett2018}
Barrett, C.W., Tinelli, C.: Satisfiability modulo theories. In: Clarke, E.M., Henzinger, T.A., Veith, H., Bloem, R. (eds.) Handbook of Model Checking., pp. 305--343. Springer (2018). \doi{10.1007/978-3-319-10575-8\_11}

\bibitem{bhayat-lpar23}
Bhayat, A., Korovin, K., Kov{\'{a}}cs, L., Schoisswohl, J.: Refining unification with abstraction. In: Piskac, R., Voronkov, A. (eds.) {LPAR} 2023: Proceedings of 24th International Conference on Logic for Programming, Artificial Intelligence and Reasoning, Manizales, Colombia, 4-9th June 2023. EPiC Series in Computing, vol.~94, pp. 36--47. EasyChair (2023). \doi{10.29007/H65J}

\bibitem{bradley-vmcai06}
Bradley, A.R., Manna, Z., Sipma, H.B.: What's decidable about arrays? In: Verification, Model Checking, and Abstract Interpretation Conference, {VMCAI}. vol.~3855, pp. 427--442. Springer (2006). \doi{10.1007/11609773\_28}

\bibitem{sc2}
Brown, C.E., Janota, M., Ol{\v{s}}{\'{a}}k, M.: Symbolic computation for all the fun. In: Brown, C.W., Kaufmann, D., Nalon, C., Steen, A., Suda, M. (eds.) Joint Proceedings of the 9th Workshop on Practical Aspects of Automated Reasoning {(PAAR)} and the 9th Satisfiability Checking and Symbolic Computation Workshop (SC-Square), 2024 co-located with the 12th International Joint Conference on Automated Reasoning {(IJCAR} 2024). {CEUR} Workshop Proceedings, vol.~3717, pp. 111--121. CEUR-WS.org (2024), \url{https://ceur-ws.org/Vol-3717/paper6.pdf}

\bibitem{Brown:04}
Brown, C.W.: {QEPCAD B}: a system for computing with semi-algebraic sets via cylindrical algebraic decomposition. SIGSAM Bull.  \textbf{38}(1),  23--24 (2004). \doi{10.1145/980175.980185}

\bibitem{burden2015numerical}
Burden, R.L., Faires, J.~Douglas, A.M.B.: Numerical Analysis. Cengage Learning, Boston, MA, 10 edn. (2015)

\bibitem{Caviness:98}
Caviness, B.F., Johnson, J.R. (eds.): Quantifier Elimination and Cylindrical Algebraic Decomposition. Springer, Wien (1998). \doi{https://doi.org/10.1007/978-3-7091-9459-1}

\bibitem{CollinsH91}
Collins, G.E., Hong, H.: Partial cylindrical algebraic decomposition for quantifier elimination. J. Symb. Comput.  \textbf{12}(3),  299--328 (1991). \doi{10.1016/S0747-7171(08)80152-6}

\bibitem{Davenport:88}
Davenport, J.H., Heintz, J.: Real quantifier elimination is doubly exponential. Journal of Symbolic Computation  \textbf{5},  29--35 (1988). \doi{10.1016/S0747-7171(88)80004-X}

\bibitem{mathconstruct}
Dekoninck, J., Balunovi{\'c}, M., Jovanovi{\'c}, N., Petrov, I., Vechev, M.: {MathConstruct}: Challenging {LLM} reasoning with constructive proofs. In: ICLR 2025 Workshop: VerifAI: AI Verification in the Wild (2025), \url{https://openreview.net/forum?id=nHW2tiGMrb}

\bibitem{DetlefsNS05}
Detlefs, D., Nelson, G., Saxe, J.B.: Simplify: {A} theorem prover for program checking. J. {ACM}  \textbf{52}(3),  365--473 (2005). \doi{10.1145/1066100.1066102}

\bibitem{ge-moura-cav09}
Ge, Y., de~Moura, L.M.: Complete instantiation for quantified formulas in satisfiabiliby modulo theories. In: Computer Aided Verification {CAV}. vol.~5643, pp. 306--320. Springer (2009). \doi{10.1007/978-3-642-02658-4\_25}

\bibitem{HoenickeSchindler-vmcai2021}
Hoenicke, J., Schindler, T.: Incremental search for conflict and unit instances of quantified formulas with {E}-matching. In: Henglein, F., Shoham, S., Vizel, Y. (eds.) Verification, Model Checking, and Abstract Interpretation - 22nd International Conference, {VMCAI} 2021, Copenhagen, Denmark, January 17-19, 2021, Proceedings. Lecture Notes in Computer Science, vol. 12597, pp. 534--555. Springer (2021). \doi{10.1007/978-3-030-67067-2\_24}

\bibitem{hozzova-cade23}
Hozzov{\'{a}}, P., Kov{\'{a}}cs, L., Norman, C., Voronkov, A.: Program synthesis in saturation. In: Pientka, B., Tinelli, C. (eds.) Automated Deduction - {CADE} 29 - 29th International Conference on Automated Deduction. LNCS, vol. 14132, pp. 307--324. Springer (2023). \doi{10.1007/978-3-031-38499-8\_18}

\bibitem{hozzova-ijcar24}
Hozzová, P., Amrollahi, D., Hajdu, M., Kovács, L., Voronkov, A., Wagner, E.M.: Synthesis of recursive programs in saturation. In: International Joint Conference on Automated Reasoning {IJCAR} (2024). \doi{10.1007/978-3-031-63498-7\_10}

\bibitem{data}
Janota, M.: Allfun (May 2025). \doi{10.5281/zenodo.15554870}

\bibitem{janota2021fair}
Janota, M., Barbosa, H., Fontaine, P., Reynolds, A.: Fair and adventurous enumeration of quantifier instantiations. In: Formal Methods in Computer-Aided Design. pp. 256--260. {IEEE} (2021). \doi{10.34727/2021/ISBN.978-3-85448-046-4\_35}

\bibitem{vampire}
Kov{\'{a}}cs, L., Voronkov, A.: First-order theorem proving and {Vampire}. In: Sharygina, N., Veith, H. (eds.) Computer Aided Verification - 25th International Conference, {CAV}. LNCS, vol.~8044, pp. 1--35. Springer (2013). \doi{10.1007/978-3-642-39799-8\_1}

\bibitem{Kuczma2009}
Kuczma, M.: An Introduction to the Theory of Functional Equations and Inequalities. Birkh\"{a}user Basel (2009). \doi{10.1007/978-3-7643-8749-5}

\bibitem{DBLP:conf/pldi/KuncakMPS10}
Kuncak, V., Mayer, M., Piskac, R., Suter, P.: Complete functional synthesis. In: {PLDI}. pp. 316--329. {ACM} (2010). \doi{10.1145/1806596.1806632}

\bibitem{sympy}
Meurer, A., Smith, C.P., Paprocki, M., {\v{C}}ert{\'\i}k, O., Kirpichev, S.B., Rocklin, M., Kumar, A., Ivanov, S., Moore, J.K., Singh, S., et~al.: {SymPy}: symbolic computing in {Python}. PeerJ Computer Science  \textbf{3}, ~e103 (2017), \url{https://www.sympy.org}

\bibitem{z3}
de~Moura, L.M., Bj{\o}rner, N.: {Z3:} an efficient {SMT} solver. In: Tools and Algorithms for the Construction and Analysis of Systems, 14th International Conference, {TACAS} 2008. vol.~4963, pp. 337--340. Springer (2008). \doi{10.1007/978-3-540-78800-3\_24}

\bibitem{musil}
Musil, V.: Funkcion\'aln\'i rovnice (2024), \url{https://prase.cz/library/FunkcionalniRovniceVM/FunkcionalniRovniceVM.pdf}, online library of the Matematick\'y koresponden\v{c}n\'i semin\'a\v{r} PraSe (PRA\v{z}sk\'y SEmin\'a\v{r}), Downloaded 8 March 2024

\bibitem{niemetz-tacas21}
Niemetz, A., Preiner, M., Reynolds, A., Barrett, C.W., Tinelli, C.: Syntax-guided quantifier instantiation. In: Tools and Algorithms for the Construction and Analysis of Systems, {TACAS}. Springer (2021). \doi{10.1007/978-3-030-72013-1\_8}

\bibitem{bluff}
Petrov, I., Dekoninck, J., Baltadzhiev, L., Drencheva, M., Minchev, K., Balunovi\'{c}, M., Jovanovi\'{c}, N., Vechev, M.: Proof or bluff? evaluating {LLMs} on 2025 usa math olympiad (2025), \url{https://arxiv.org/abs/2503.21934}

\bibitem{ratschan-mfcs23}
Ratschan, S.: Deciding predicate logical theories of real-valued functions. In: Leroux, J., Lombardy, S., Peleg, D. (eds.) 48th International Symposium on Mathematical Foundations of Computer Science ({MFCS} 2023). LIPIcs, vol.~272, pp. 76:1--76:15. Schloss Dagstuhl -- Leibniz-Zentrum f{\"u}r Informatik, Dagstuhl, Germany (2023). \doi{10.4230/LIPIcs.MFCS.2023.76}

\bibitem{reger-tacas18}
Reger, G., Suda, M., Voronkov, A.: Unification with abstraction and theory instantiation in saturation-based reasoning. In: Beyer, D., Huisman, M. (eds.) Tools and Algorithms for the Construction and Analysis of Systems - 24th International Conference, {TACAS} 2018, Held as Part of the European Joint Conferences on Theory and Practice of Software, {ETAPS} 2018, Thessaloniki, Greece, April 14-20, 2018, Proceedings, Part {I}. Lecture Notes in Computer Science, vol. 10805, pp. 3--22. Springer (2018). \doi{10.1007/978-3-319-89960-2\_1}

\bibitem{ReynoldsBF18}
Reynolds, A., Barbosa, H., Fontaine, P.: Revisiting enumerative instantiation. In: Tools and Algorithms for the Construction and Analysis of Systems (TACAS). Springer (2018). \doi{10.1007/978-3-319-89963-3\_7}

\bibitem{reynolds-cade13}
Reynolds, A., Tinelli, C., Goel, A., Krsti\'c, S., Deters, M., Barrett, C.: Quantifier instantiation techniques for finite model finding in {SMT}. In: 24th International Conference on Automated Deduction, CADE 2013. pp. 377--391 (2013). \doi{10.1007/978-3-642-38574-2\_26}

\bibitem{reynolds-fmcad14}
Reynolds, A., Tinelli, C., de~Moura, L.M.: Finding conflicting instances of quantified formulas in {SMT}. In: Formal Methods in Computer-Aided Design, {FMCAD}. pp. 195--202. {IEEE} (2014). \doi{10.1109/FMCAD.2014.6987613}

\bibitem{handbookAR}
Robinson, J.A., Voronkov, A. (eds.): Handbook of Automated Reasoning (in 2 volumes). Elsevier and {MIT} Press (2001), \url{https://www.sciencedirect.com/book/9780444508133/handbook-of-automated-reasoning}

\bibitem{srivastava13}
Srivastava, S., Gulwani, S., Foster, J.S.: Template-based program verification and program synthesis. Int. J. Softw. Tools Technol. Transf.  \textbf{15}(5-6),  497--518 (2013). \doi{10.1007/S10009-012-0223-4}

\bibitem{Tarski:51}
Tarski, A.: A Decision Method for Elementary Algebra and Geometry. Univ. of California Press, Berkeley (1951). \doi{https://doi.org/10.2307/jj.8501420}, also in~\cite{Caviness:98}

\bibitem{tarski}
Vale{-}Enriquez, F., Brown, C.W.: Polynomial constraints and unsat cores in {Tarski}. In: Davenport, J.H., Kauers, M., Labahn, G., Urban, J. (eds.) Mathematical Software - {ICMS} - 6th International Conference. LNCS, vol. 10931, pp. 466--474. Springer (2018). \doi{10.1007/978-3-319-96418-8\_55}, \url{https://www.usna.edu/Users/cs/wcbrown/tarski/index.html}, code obtained from \url{https://github.com/chriswestbrown/tarski}.

\bibitem{Weispfenning:88}
Weispfenning, V.: The complexity of linear problems in fields. Journal of Symbolic Computation  \textbf{5}(1--2),  3--27 (1988). \doi{10.1016/S0747-7171(88)80003-8}

\end{thebibliography}

\end{document}